\documentstyle[twoside,fleqn,espcrc2,psfig]{article}
\newcommand{\be}{\begin{eqnarray}}
\newcommand{\ee}{\end{eqnarray}}
\newcommand{\bea}{\begin{eqnarray}}
\newcommand{\eea}{\end{eqnarray}}
\newcommand{\ba}{\begin{array}}
\newcommand{\ea}{\end{array}}

\newcommand{\no}{\nonumber}

\newcommand{\tr}{\triangle}
\newcommand{\trv}{\vec{\triangle}}

\newcommand{\x}{{\mbox{$\vec{x}$}}}

\newcommand{\sig}{{\mbox{$\vec{\sigma}$}}}
\newcommand{\B}{{\mbox{$\vec{B}$}}}
\newcommand{\E}{{\mbox{$\vec{E}$}}}
\newcommand{\gam}{{\mbox{$\vec{\gamma}$}}}
\newcommand{\Sig}{{\mbox{$\vec{\Sigma}$}}}

\def\simgt{\rlap{\lower 3.5 pt\hbox{$\mathchar \sim$}}
           \raise 1pt \hbox {$>$}}
\def\simlt{\rlap{\lower 3.5 pt\hbox{$\mathchar \sim$}}
           \raise 1pt \hbox {$<$}}

\newcommand{\AmS}{{\protect\the\textfont2
  A\kern-.1667em\lower.5ex\hbox{M}\kern-.125emS}}
\title{A study of $O(1/m_{Q}^{2})$ corrections for $f_{B}$ 
       with lattice NRQCD}
\author{
N. Yamada\address{Department of Physics, Hiroshima University,
Higashi-Hiroshima 739, Japan}\thanks{Presented by N.Yamada.
This work was partly supported by Monbusho International
Scientific Reserach Program (No. 08044089).},
S. Hashimoto\address{High Energy Accelerator Research
Organization(KEK), Tsukuba 305,
Japan}\thanks{S.H. is supported by Ministry of Education, Science and
Culture under grant number 09740226.},
K-I. Ishikawa$^{\rm a}$,
H. Matsufuru$^{\rm a}$\thanks{H.M. would like to thank
the JSPS for Young Scientists for a research fellowship.}
and T. Onogi$^{\rm a}$
}
\begin{document}
\begin{abstract}
\vspace{-0.2cm}
We investigate higher order effects
in the nonrelativistic expansion of lattice QCD
on the heavy-light meson decay constants and some other quantities 
in order to understand the truncation error of NRQCD.
While our numerical results have
large $O(a)$ and $O(\alpha_s)$ errors
due to the use of Wilson light quark action
and the tree-level matching,
we find that the truncation error of higher order
relativistic corrections are adequately small
around the mass of the $b$ quark. 
We also present a perturbatively matched results through 1-loop level
without operator mixing effects.
\vspace{-0.5cm}
\end{abstract}
\maketitle
\section{Introduction}
\vspace{-0.2cm}
It is indispensable
for the verification of the Standard Model and
the new physics search
that weak matrix elements between hadrons involving $b$-quark
are calculated with high accuracy.
Here, we focus on the heavy-light meson decay constants\cite{matsu},
particularly $f_B$, whose accurate value is needed
for the determination of $|V_{tb}^*V_{td}|$
together with $B_B$ parameter.
The lattice NRQCD\cite{nrqcd} enable us
to do the direct simulations at the $b$ quark
and has yielded remarkable progress
on the heavy quark physics.
So far the studies including $1/m_Q$ corrections
have made it clear
that there exists a large $1/m_Q$ correction
to the value in the static limit\cite{SGO}.
In this talk we report on the systematic study
of the $1/m_Q^2$ corrections.
A preliminary report of an investigation
of $O(1/m_Q^2)$ corrections similar to our work
has been reported in Ref.\cite{fbm2}.
The action and operators should be matched
to the full theory
since NRQCD is an effective theory.
Recently we have completed 
the self-energy renormalization constants
and the multiplicative part in the renormalization
of axial vector current\cite{ishi}.
In this article we present the results of $f_B$
and some other quantities with and without 1-loop matching.

\vspace{-0.2cm}
\section{Simulations}
\vspace{-0.2cm}
Our numerical simulation is carried out
with 120 quenched configurations
on a $16^3 \times 32$ lattice at $\beta = 5.8$.  
For the tadpole factor
we employ $u_0 = \langle P_{plaq} \rangle^{1/4}$
with $P_{plaq}$ the average plaquette,
which takes the value $u_0 = 0.867994(13)$
measured during our configuration generation.

For the light quark we use the Wilson quark action
with $\kappa$=0.1570, 0.1585 and 0.1600. 
The chiral limit is reached  at $\kappa_c = 0.16346(7)$
and the inverse lattice spacing determined from the rho meson mass
equals $a^{-1} = 1.714(63)$ GeV.
The hopping parameter $\kappa_s$ corresponding to the strange quark is
determined from $m_{\phi}/m_{\rho}$ and $m_{K}/m_{\rho}$,
which yields $\kappa_s = 0.15922(39)$ and $0.16016(23)$, respectively.
We use the factor $\sqrt{1-\frac{3\kappa}{4\kappa_c}}$
as the field normalization for light quark\cite{KLM}.

For the heavy quark, we use the lattice NRQCD
with the following evolution equation:
\be
&& \hspace{-0.7cm}
G_Q(t\ge0,\x) = 
   \left( 1 - \frac{aH_0}{2n} \right)^n 
   \left( 1 - \frac{a\delta H}{2} \right) 
  U_4^{\dag} \no \\
&& 
\times  \left( 1 - \frac{a\delta H}{2} \right) 
   \left( 1 - \frac{aH_0}{2n} \right)^n 
   G_Q(t-1,\x)                         \no \\ 
&& 
+  \delta_{x,0}.
\ee
Where $H_0$ is defined as follows:
\be
H_0 = - \frac{\triangle^{(2)}}{2m_Q},
\ee
and in order to realize two accuracies
we define two $\delta H$'s as follows:
\be
    \delta H_1
&=& - \frac{g\sig\cdot\B}{2m_Q},\\
      \delta H_2
&=& - \frac{g\sig\cdot\B}{2m_Q}
    + \frac{ig}{8m_Q^2}( \trv \cdot \E - \E \cdot \trv ) \no \\
& & - \frac{g}{8m_Q^2}\sig \cdot(\trv \times \E - \E \times \trv ) \no \\
& & - \frac{(\triangle^{(2)})^2}{8m_Q^3}
    +  \frac{a^2 \tr^{(4)}}{24m_Q}
    - \frac{a (\tr^{(2)})^2}{16nm_Q^2}.
\ee
The symbols $\trv$ and $\tr^{(2)}$ denote the symmetric lattice
differentiation in spatial directions and Laplacian,
respectively, and $\tr^{(4)} \equiv \sum_i (\tr^{(2)}_i)^2$.
The field strengths $\B$ and $\E$ are generated from the standard 
clover-leaf operator.

The original 4-component heavy quark spinor $h$
is related to two 2-component spinors $Q$ and $\chi$
through Foldy-Wouthuysen-Tani(FWT) transformation,
\be
h(x) = R \left( \begin{array}{c} Q(x) \\ \chi^{\dag}(x) 
                    \end{array} \right),
\ee
where $R$ is an inverse FWT transformation matrix which has
$4\times4$ spin and $3\times3$ color indices.
We prepare two $R$'s as before.
After discretization, $R$ at the tree level is written as
follows:
\be
&&\hspace{-0.7cm}
    R_1
= 1 - \frac{\gam\cdot\trv}{2m_Q}, \\
&&\hspace{-0.7cm}
    R_2
= 1 - \frac{\gam\cdot\trv}{2m_Q} 
    + \frac{\tr^{(2)}}{8m_Q^2}
    + \frac{g\Sig\cdot\B}{8m_Q^2}   
    - \frac{ig\gamma_4\gam\cdot\E}{4m_Q^2},   
\ee
where $\Sigma^{j} = {\rm diag.}\{ \sigma^{j}, \sigma^{j} \}$.
We apply the tadpole improvement\cite{MFimp}
to all link variables in the evolution equation and $R$
by rescaling the link variables
as $U_{\mu} \rightarrow U_{\mu}/u_0$.

We performed two simulations using following two sets: 
\be 
\mbox{set\ I} \equiv \{ \delta H_1,R_1 \}
\hspace{0.3cm}{\rm and}\hspace{0.3cm} 
\mbox{set\ I$\!$I} \equiv \{ \delta H_2,R_2 \}.
\ee
set I and I$\!$I are fully consistent
through $O(1/m_Q)$ and $O(1/m_Q^2)$, respectively.
Furthermore $\delta H_2$ has the leading relativistic correction
to the dispersion relation, which is an $O(1/m_Q^3)$ term,
and the terms improving the discretization errors
appearing in $H_0$ and time evolution are also included.
For the heavy quark mass and the stabilizing parameter,
we use $(am_Q,n)$=(10.0,1), (5.0,2), (2.6,2), (2.1,3), (1.5,3),
(1.2,3) and (0.9,4), which cover a mass range between $4m_b$ and $m_c$. 

Our strategy is to study systematically the effects of
an inclusion of $O(1/m_Q^2)$ terms
by comparing two results from these simulations.
All of our errors are estimated by a single elimination jack-knife
procedure. 

\vspace{-0.2cm}
\section{Results}
\vspace{-0.2cm}
\subsection{decay constants}
\begin{figure}[tb]
\vspace{12pt}
\leavevmode\psfig{file=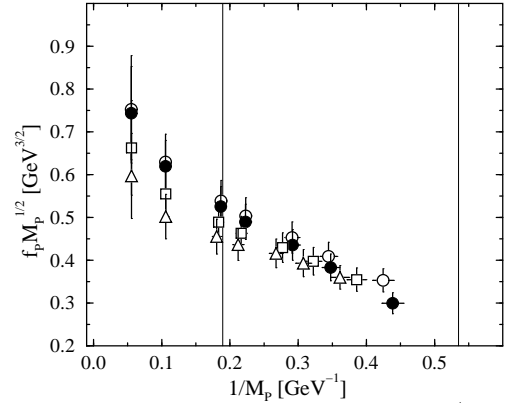,width=6.5cm}
\vspace{-1.2cm}
\caption{$1/M_P$ dependence of $f_P{M_P}^{1/2}$
with set I(open symbols) and set I$\!$I(solid symbols).
Circles are the results with tree-level matching
and squares and triangles are with 1-loop matching
on the scale of $q^*$=$\pi/a$ and $1/a$, respectively.}
\label{fig1}
\vspace{-0.8cm}
\end{figure}

Fig.\ref{fig1} shows the $1/M_P$ dependence of $f_P\sqrt{M_P}$.
Where $M_P$ is a pseudoscalar heavy-light meson mass.
Two vertical lines represent the mass regions
corresponding to the B and D meson.
Circles(open for set I and filled for set I$\!$I)
are the results with tree-level matching.
From more quantitative study, we can find
that within tree-level analysis
the relative magnitude of $(1/m_Q^2)$ correction is
about 3\%, conservatively 6\%, around the $B$ meson region\cite{our}.

\vspace{-0.2cm}
\subsection{other quantities}
\begin{figure}[t]
\vspace{12pt}
\leavevmode\psfig{file=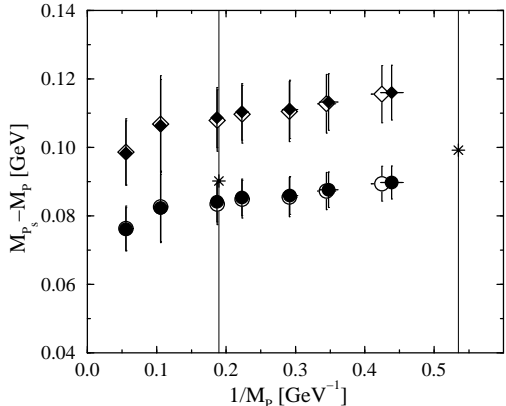,width=6.5cm}
\vspace{-1.2cm}
\caption{$1/M_P$ dependence of $M_{P_s}-M_P$ with set I(open symbols)
and set II(solid symbols).
Diamonds are with $\kappa_s$ from $\phi$ meson
and circles from $K$ meson.
Two asterisks indicate the experimental values.}
\label{fig2}
\vspace{-0.8cm}
\end{figure}

\begin{figure}[t]
\vspace{12pt}
\leavevmode\psfig{file=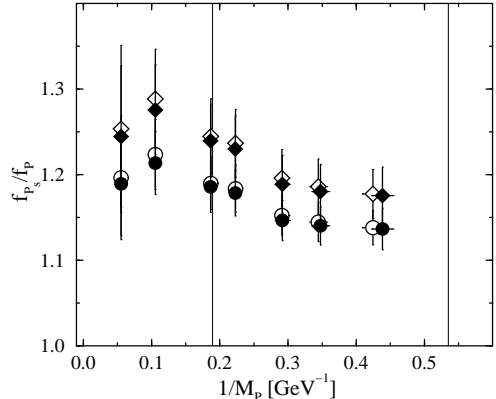,width=6.5cm}
\vspace{-1.2cm}
\caption{$1/M_P$ dependence of  $f_{P_s}/f_P$ with set I(open symbols)
and set II(solid symbols). 
Diamonds are with $\kappa_s$ from $\phi$ meson
and circles from $K$ meson.}
\label{fig3}
\vspace{-0.8cm}
\end{figure}

We also performed a similar investigation
on $M_{P_s} - M_P$ and $f_{P_s}/f_P$,
where $P_s$ denotes a pseudoscalar heavy-strange meson.
Those results are shown in Fig.\ref{fig2} and \ref{fig3}.
One can see from these figures
that there is no significant difference
between the results from the two sets
over almost all mass region.
\vspace{-0.2cm}
\subsection{perturbative matching}
Recently we have completed 
the self-energy renormalization constants
and the multiplicative part in the renormalization
of axial vector current with set I\cite{ishi}.
Since the scale $q^*$
where the strong coupling constant should be estimated
is still undetermined, we estimate the matching factor
at $q^*=1/a$ and $\pi/a$ and show the results
with the factor in Fig.\ref{fig1}.

\vspace{-0.2cm}
\section{Summary}
\vspace{-0.2cm}
We presented the effects of $O(1/m_Q^2)$ correction
to the heavy-light meson decay constant
with tree-level matched lattice NRQCD and Wilson quark action
in a quenched approximation.
While the $O(1/m_Q)$ correction to the decay constant in the
static limit is significant\cite{SGO}, we find in our
systematic study that the $O(1/m_Q^2)$ correction is
sufficiently small for B meson, so that there will be no
need for incorporating $O(1/m_Q^3)$ corrections unless an
accuracy of better than 5\% is sought for.
Our examination of other physical quantities in the same
respect also provides encouraging support to this statement. 
We have thus shown using our highly improved lattice NRQCD
that the relativistic error, which has been one of the largest
uncertainties in lattice calculations of the B meson decay
constant, is well under control. 
Our results still have several sources of large systematic
errors, that is, $O(a)$, $O(\alpha_s/M)$, $O(a\alpha_s)$ and
a quenching errors.
When improvements of these errors are all in place,
we expect to achieve a lattice NRQCD determination of
$f_B$ with the accuracy of less than 10\%.

Numerical calculations have been done on Paragon XP/S at
INSAM (Institute for Numerical Simulations and Applied
Mathematics) in Hiroshima University.
\vspace{-0.2cm}

%
\end{document}